\documentstyle[psfig]{l-aa}

\begin{document}

\thesaurus {06 (04.19.11, 08.02.2, 08.02.4, 08.05.1, 08.06.3)}

\title{Evaluating GAIA performances on eclipsing binaries.}
\subtitle{I. Orbits and stellar parameters for V505 Per, V570 Per and OO Peg}

\author{
       U. Munari\inst{1,2}
\and   T. Tomov\inst{1,3}
\and   T. Zwitter\inst{4}
\and   J. Kallrath\inst{5,6}
\and   E.F. Milone\inst{7}
\and   P.M. Marrese\inst{1,2}
\and   F. Boschi\inst{1,8}
\and\\ A. Pr\v sa\inst{4}
\and   L. Tomasella\inst{1}
\and   D. Moro\inst{9} 
       }
\offprints{U.Munari}

\institute {
Osservatorio Astronomico di Padova, Sede di Asiago, I-36012 Asiago (VI), Italy
\and 
CISAS, Centro Interdipartimentale Studi ed Attivit\`a Spaziali dell'Universit\`a di Padova, Italy
\and
Centre for Astronomy, Nicholaus Copernicus University, ul. Gagarina 11, 87-100 Torun, Poland
\and
University of Ljubljana, Department of Physics, Jadranska 19, 1000 Ljubljana, Slovenia 
\and
BASF-AG, ZDP/C-C13, D-67056 Ludwigshafen, Germany
\and
Astronomy Department, University of Florida, Gainesville, FL 32611, USA
\and
Physics and Astronomy Department, University of Calgary, Calgary T2N 1N4, Canada
\and
Dipartimento di Fisica dell'Universit\`a di Milano, via Celoria 20, 20131 Milano, Italy
\and
Dipartimento di Astronomia dell'Universit\`a di Padova, Osservatorio Astrofisico, I-36012 Asiago (VI), Italy
}
\date{Received date..............; accepted date................}

\maketitle

\begin{abstract}
The orbits and physical parameters of three detached, double-lined A-F
eclipsing binaries have been derived combining $H_P, V_T, B_T$ photometry
from the Hipparcos/Tycho mission with 8500-8750 \AA\ ground-based
spectroscopy, mimicking the photometric+spectroscopic observations that
should be obtained by GAIA, the approved Cornerstone 6 mission by ESA. This
study has two main objectives, namely ($a$) to derive reasonable orbits for
a number of new eclipsing binaries and ($b$) to evaluate the expected
performances by GAIA on eclipsing binaries and the accuracy achievable on
the determination of fundamental stellar parameters like masses and radii.
It is shown that a 1\% precision in the basic stellar parameters can be
achieved by GAIA on well observed detached eclipsing binaries provided that
the spectroscopic observations are performed at high enough resolution.
Other types of eclipsing binaries (including semi-detached and contact
types) and different spectral types will be investigated in following papers
along this series.
\end{abstract}
\maketitle

\section{Introduction}

\begin{table*}[!Ht]
\tabcolsep 0.08truecm
\caption{Program eclipsing binaries. Data from the Hipparcos Catalogue. $H_P, B_T, V_T$ are median values (i.e. out-of-eclipse).}
\begin{tabular}{llccccccccccc} \hline
&&&&&&&&&&&\\
Name & & Spct. & $H_P$ & $B_T$ & $V_T$ & $\alpha_{J2000}$ & $\delta_{J2000}$ & parallax & $\mu_\alpha$ & $\mu_\delta$ \\ 
     & &       &       &       &       & (h m s)          & ($^\circ$ ' ") & (mas) & (mas yr$^{-1}$) & (mas yr$^{-1}$)\\
&&&&&&&&&&&\\ \hline
&&&&&&&&&&&\\
V505 Per & HIC 10961  & F5 & 6.960 & 7.364 & 6.916 & 02 21 12.9625 & +54 30 36.282 &  15.00$\pm$0.84 & 41.61$\pm$0.63 &~~--3.74$\pm$0.47 \\
V570 Per & HIC 14673  & F5 & 8.170 & 8.641 & 8.128 & 03 09 34.9443 & +48 37 28.696 & ~~8.53$\pm$0.96 & 52.20$\pm$0.85 & --41.58$\pm$0.79 \\
OO Peg   & HIC 107099 & A2 & 8.336 & 8.690 & 8.359 & 21 41 37.6982 & +14 39 30.747 & ~~2.24$\pm$1.05 &--3.11$\pm$1.29 & --14.37$\pm$0.71 \\
&&&&&&&&&&&\\
\hline
\end{tabular}
\end{table*}

Eclipsing binaries are a prime tool to derive fundamental stellar
parameters like mass and radius, or the temperature scale. The study of
eclipsing binaries is by no means a simple task as evidenced by the fact
that for less than a hundred objects the stellar parameters have been
derived with an accuracy of 1\% or better. The prospects for the future
sound however quite bright given the expected performances of the GAIA
mission.

GAIA has just been selected as the next ESA Cornerstone~6 mission and it is
designed to obtain extremely precise astrometry (in the {\sl micro-}arcsec
regime), multi-band photometry and medium/high resolution spectroscopy for a
large sample of stars.  The goals as depicted in the mission {\sl Concept
and Technology Study} (ESA SP-2000-4) call for astrometry and broad band
photometry to be collected for all stars down to $V \sim$20 mag over the
entire sky 
($\sim 1\cdot 10^9$ stars), with brighter magnitude limits for
spectroscopy and intermediate band photometry. Each target star should be
measured over a hundred times during the five year mission life-time, in a
fashion similar to the highly successful {\sl Hipparcos} operational mode.
The astrophysical guidelines of the GAIA mission are discussed by Gilmore et
al. (1998) and Perryman et al. (2001), an overview of the GAIA payload and
spacecraft being presented by M\'erat et al. (1999), while the goals of 
GAIA spectroscopy and photometry have been discussed by Munari (1999a,b).

With photometry complete down to $V=20$ mag and spectroscopy down to $V=15$
mag or so (depending on the final optical design and overall throughput),
one may expect to detect a large number of eclipsing binaries. Let's roughly
estimate how much. The number of stars brighter than $V=15$ (thus bright
enough for GAIA to obtain {\sl both} photometry {\sl and} spectroscopy)
is 5$\cdot 10^7$ and their average spectral type is about G7. Scaling the
Hipparcos results (917 detected among the 118218 stars surveyed, or 0.8\%),
$\sim4\cdot 10^5$ of them should be eclipsing binaries. At an average G7
spectral type it may be estimated that about 25\% of them ( $\sim1\cdot 10^5$)
will be double-lined in GAIA spectral observations (cf. Carquillat et al.
1982).  Even if for only 1\% of them it should be possible to derive orbits
and stellar parameters at 1\% precision, this still would be $\sim25\times$
what have been so far collected from devoted ground-based observing
campaigns during the last century (cf. Andersen 1991).

GAIA is baselined to rotate every three hours around an axis (pointing
55$^\circ$ away from the Sun) that completes a precession cycle every 76
days. Stars will be recorded while they transit across the field of view of
the three fixed telescopes on board (which use as detectors mosaics of CCDs
operating in time-delayed integration mode). Thus, how many times and when a
star will be observed depends on its position on the sky and the GAIA
scanning law. On the average a star should be observed $\sim$150 times
during the GAIA mission lifetime. On each passage over a given star, GAIA
will obtain one astrometric measurement, one reading each for a series of
photometric bands ($\sim$11 are currently baselined) and one spectrum (covering
the region 8500-8750 \AA\, centered on the near infrared Ca~II triplet and
the head of the Paschen series).

\begin{table}[!Hb]
\tabcolsep 0.08truecm
\caption{Number of Hipparcos ($H_P$) and Tycho ($B_T$, $V_T$) photometric data 
and ground based radial velocity observations, their mean S/N and standard error
for the three program stars.}
\begin{tabular}{lccccccccccc} \hline
&&
\multicolumn{2}{c}{\sl Hip}&&
\multicolumn{3}{c}{\sl Tyc}&&
\multicolumn{3}{c}{\sl RV}\\ \cline{3-4} \cline{6-8} \cline{10-12} 
\multicolumn{11}{c}{}\\
&&
N&$\sigma$($H_P$)&&
N&$\sigma$($B_T$)&$\sigma$($V_T$)&&
N&S/N&$\sigma$({\sl RV})\\
  OO Peg &&  73 & 0.013 &&  61 & 0.110 & 0.109 && 21 & 38 & 3.5 \\
V570 Per &&  92 & 0.014 && 120 & 0.111 & 0.100 && 28 & 57 & 3.6 \\
V505 Per && 122 & 0.009 && 152 & 0.057 & 0.051 && 20 & 91 & 1.7 \\  
\hline
\end{tabular}
\end{table}

\begin{figure}[!Hb]
\centerline{\psfig{file=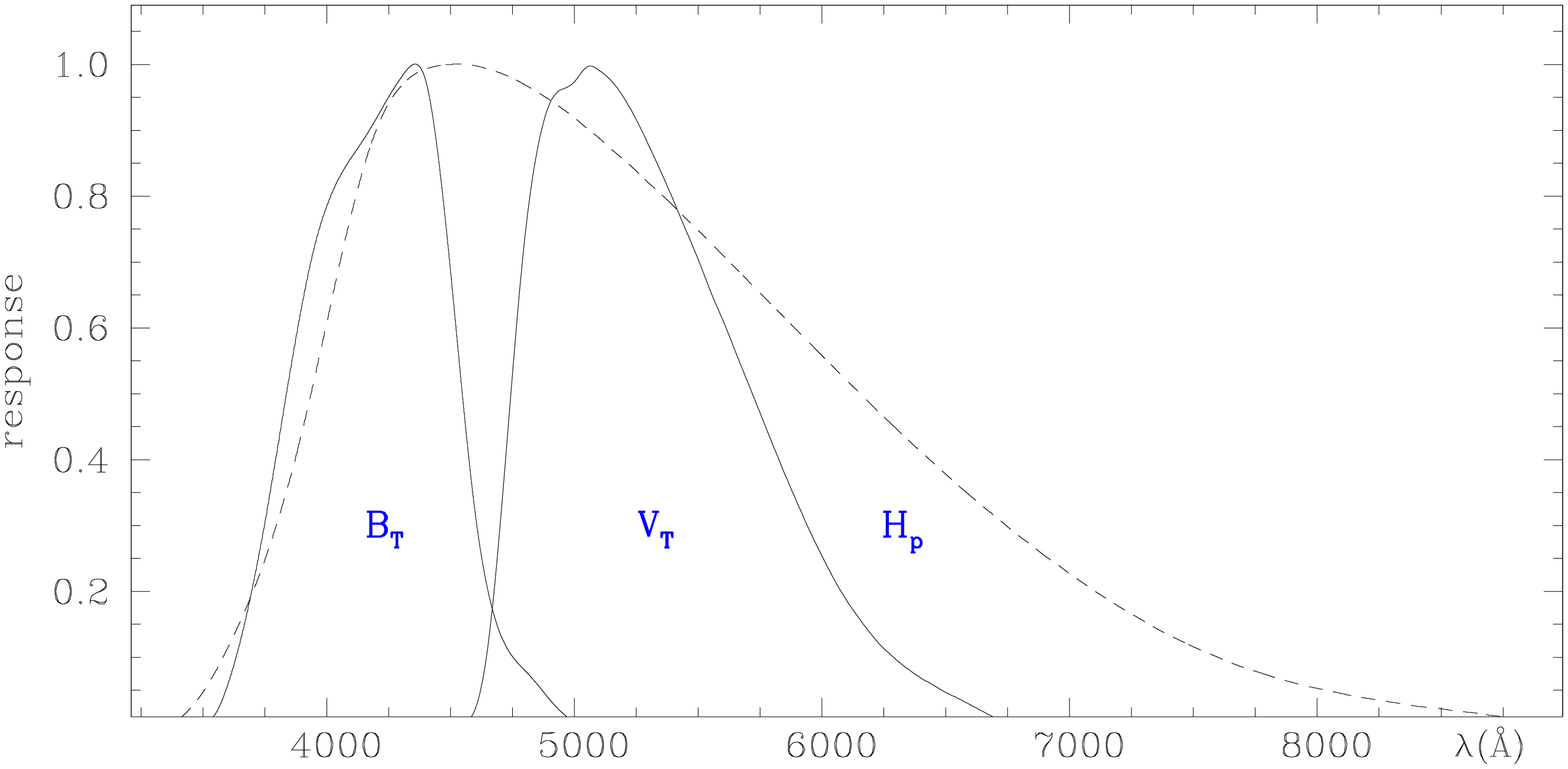,width=8.9cm,angle=270}}
\caption[]{Transmission of the Hipparcos $H_P$
and Tycho $B_T, V_T$ bands.}
\end{figure}

How this sampling by GAIA compares with examples of state-of-the-art ground
based studies of eclipsing binary stars~? Generally, between 30 and 60 good
spectra (secured away from conjunction phases) are enough to 
characterize well the radial velocity curve and the spectroscopic orbit, while
many hundreds to several thousands photometric points are necessary to cover
in detail the photometric lightcurve (and the eclipse phases in particular)
and to derive an accurate photometric solution. The 150 spectra per star
secured on the average by GAIA should put the spectral monitoring of
eclipsing binaries on the safe side even for the faintest recorded systems.
Where GAIA could fall short of optimal coverage is with the photometric
observations. A pertinent example is given by Hipparcos, which scanned the
sky in a fashion very similar to GAIA: with a mean of 110 observations per
star, the eclipses of many binaries have been covered by less than 10
photometric points.

The limit on the number of photometric observations is intrinsic to the GAIA
operation mode and cannot be changed by large margins. It is also quite
evident that a devoted effort from an international consortium to perform
follow-up observations of even a small subsample of the $\sim4\cdot 10^5$
eclipsing binaries detected by GAIA would be an unrealistic goal to achieve. 
Moreover, it can be anticipated that such follow-up projects could take many
years to be completed and could obviously begin only after the end of the
GAIA mission and the completion of the data reduction and dissemination
phase. It thus seems quite relevant to investigate the accuracy to which
eclipsing binaries can be investigated on the base of the GAIA data alone.
The present series of papers is devoted precisely to this aim, combining
ground based spectroscopy over the 8500-8750 \AA\ interval chosen for GAIA
with existing Hipparcos/Tycho photometry that well mimics the GAIA
photometric harvest.

To be precise, GAIA is expected to collect a number of photometric points
per star, per band  similar to Hipparcos. However Hipparcos collected data in
just three bands ($H_P$, $B_T$, $V_T$, see Figure~1), while GAIA is
baselined to operate $\sim$11 bands. Therefore, combining all data in all
bands, GAIA should provide $\sim 4 \times$ photometric points per star
compared to Hipparcos.

\begin{table*}[!Ht]
\tabcolsep 0.08truecm
\caption{Journal of radial velocity data. The columns give the spectrum
number (as from the Asiago 1.82 m Echelle+CCD log book), the heliocentric JD
(+2451000), the radial velocities and their standard errors for both components.
An asterisk marks the spectra with too severe blending of the lines from both 
components for meaningful separate measurement. The latter have not been used 
in modeling the binaries.}
\centerline{\psfig{file=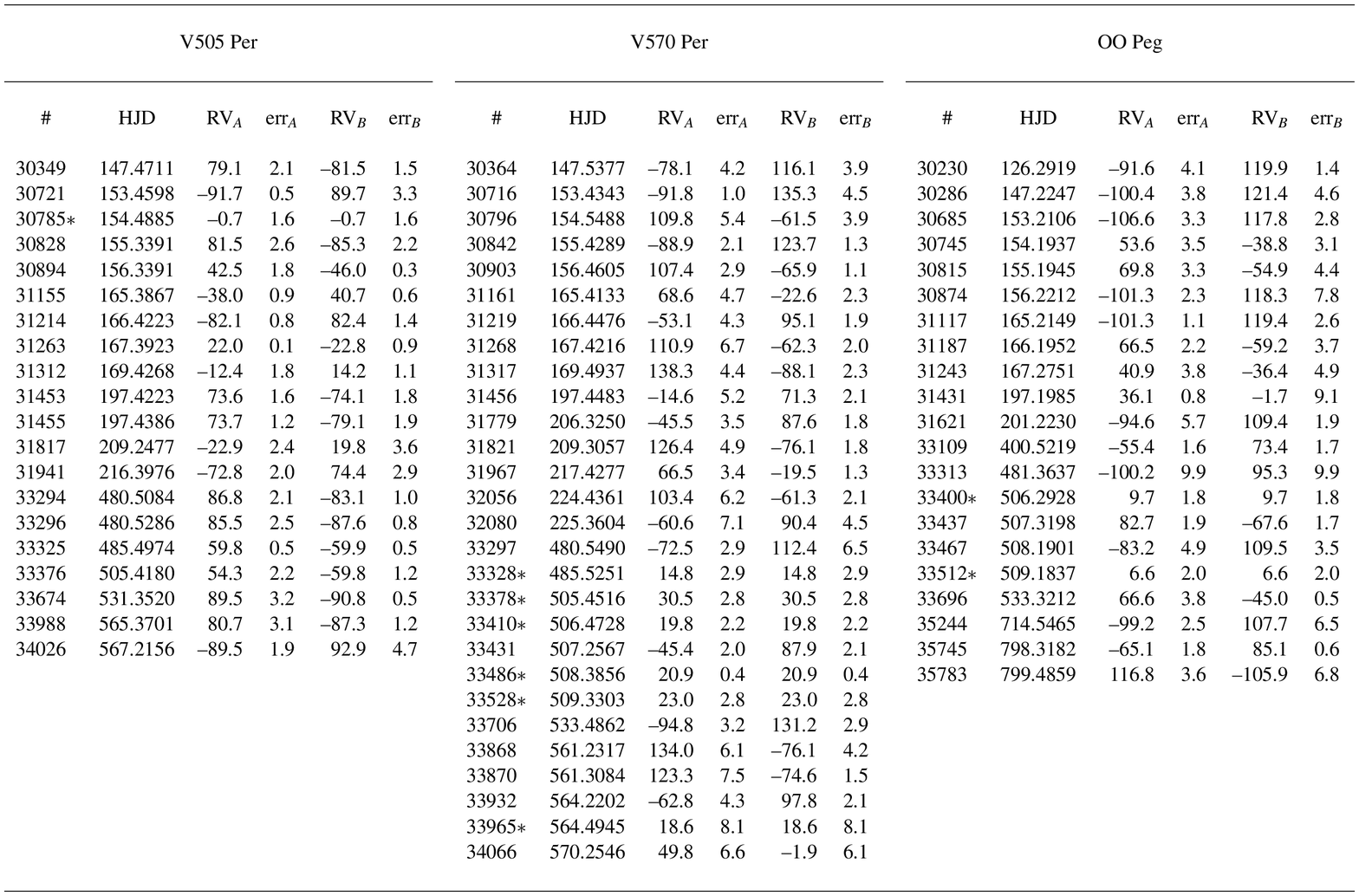,width=18cm}}
\end{table*}

The dispersion currently baselined for GAIA spectroscopy is 0.75 \AA/pix.
However, a different dispersion could eventually be selected between the
0.25 $-$ 1.5 \AA/pix boundaries investigated during the mission planning
phase. We decided to carry out our observations at the highest resolution
considered possible for GAIA (0.25 \AA/pix), for two basic reasons: ($a$) to
evaluate the maximum GAIA potential, and ($b$) because data secured at a
higher resolution can always be degraded to mimic lower resolution
observations, while the contrary is obviously not feasible.

Finally, it is worth noticing that GAIA will be able to follow the
astrometric motion around the baricenter of the components of binary stars
separated by at least 0.05 arcsec, thus providing a completely independent
method to derive fundamental stellar parameters for those binaries 
in the solar neighbourhood whose orbital period is not much longer than 
the GAIA mission lifetime.

\section{Target Selection}

At the relatively bright magnitude limit reached by Tycho observations
(limit of completeness $V\sim$10.5) the average color of the stars
corresponds to a G0 spectral type.  At the fainter magnitudes reached by
GAIA the stars will be increasingly redder, with an average K0 spectral type
at V=15 mag. Therefore the target stars for the present series of papers
will be biased toward G-K spectral types, with less target stars among the
A, F and M types.

All targets in this paper series are selected among the eclipsing binaries
observed by the Hipparcos/Tycho mission, with preference to those lacking an
orbital solution in literature. 

In this first paper we investigate the three A--F, detached eclipsing
binaries listed in Table~1. The number of available photometric and
spectroscopic data and their mean standard errors are given in Table~2. 
V505~Per has an accurate spectroscopic/photometric solution from extensive
ground observations (Marshall et al. 1997) and serves for comparison, while
for V570~Per and OO~Peg the orbit is derived here for the first time.

In following papers along this series we will also consider semi-detached
and contact eclipsing binaries as well as eclipsing binaries with one or
both of the components being a variable star itself.

\begin{figure*}[!Ht]
\centerline{\psfig{file=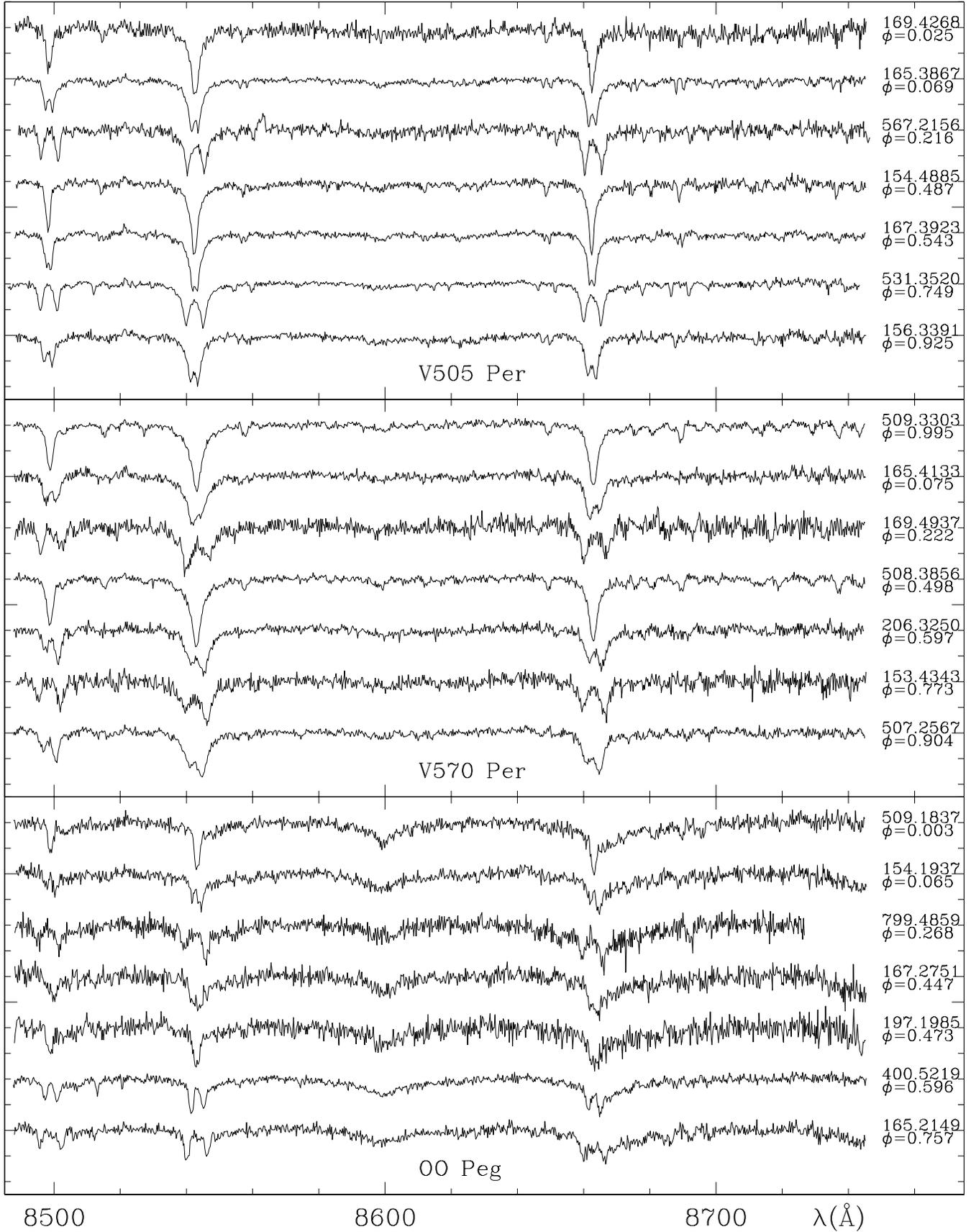,width=18cm}}
\caption[]{A sample of the recorded spectra for the program stars. The HJD
(+2451000) and the orbital phase (according to orbital elements listed 
in Table 4) are given on the right.}
\end{figure*}

\section{Radial Velocities}

A 0.25 \AA/pix dispersion and a $\sim$0.50 \AA\ resolution over the
8500-8750 \AA\ wavelength range (therefore a resolving power
R~=~$\lambda/\bigtriangleup\lambda$~=~17,000) will be maintained
throughout this paper series.

The spectroscopic observations have been obtained with the Echelle+CCD
spectrograph on the 1.82 m telescope operated by Osservatorio Astronomico di
Padova atop Mt. Ekar (Asiago). A 2.2 arcsec slit width was adopted to match
the R~=~$\lambda/\bigtriangleup\lambda$~=~17,000 requirement. The detector
has been a UV coated Thompson CCD 1024x1024 pixel, 19 micron square 
size. The GAIA spectral range is covered without gaps in a single order by
the Asiago Echelle spectrograph. The actual observations however extended
over a much larger wavelength interval (4500-9000 \AA). Here we will limit
the analysis to the GAIA spectral interval; the remaining, much larger
wavelength domain will be analyzed elsewhere together with devoted multi
band photometry from pointed ground based observations.

A sub-sample of the spectra collected at representative phases is presented
in Figure~2 for the three program stars.

The spectra have been extracted and calibrated in a standard fashion using
the IRAF software package running on a PC under Linux operating system. The
stability of the wavelength scale of the Asiago Echelle spectrograph has
been discussed in detail by Munari and Lattanzi (1992) and Munari and
Tomasella (1999a). We checked it on each recorded spectrum by measuring the
wavelengths of several telluric absorption lines that are abundant in the
Echelle orders next to the one covering the GAIA spectral window (cf. Munari
1999b). No wavelength shift has been detected in excess of 0.2 km sec$^{-1}$
which nicely compares with the intrinsic precision limit of radial
velocities from cross-correlation techniques ($\pm$0.01 pixels = 0.1 km
sec$^{-1}$ for the chosen instrumental set-up).

The radial velocities on the spectra of the three targets have been measured
both in a line-by-line fashion as well as via cross-correlation. 

In the line-by-line approach we considered for all the program binaries only
the three CaII lines (8498.018, 8542.089, 8662.140 \AA; Paschen 14  at 8598.394
\AA\ has been used too, but only on the best OO~Peg spectra). To measure the line
wavelengths we tested different approaches: profile fitting (gaussian,
lorentzian or voigt profiles), bisector method and line photo-centering
({\tt center1d} ~tool in IRAF).  They gave nearly equivalent results, with the
line-centering performing slightly better on average.

The superior intrinsic potential of the cross-correlation approach ({\tt
fxcor}\ and {\tt xcsao}\ tools in IRAF) is generally only marginally exploited
in real, ground-based spectra because of the need for a normalization of the
continuum spectrum before running the cross-correlation. The normalization is
particularly critical for our spectra (A and F spectral types, cf. Table~1)
because they are dominated by a few broad lines with extended wings and
nearly line-empty continuum in between. The shape of the continuum is badly
affected by the blaze function (particularly peaked in Echelle low-number
and near-IR orders) and by less predictable effects arising at the
spectrograph slit and condensations on the dewar window. 
As templates for the cross-correlation we both used single
lined spectra at conjunctions, as well as real and synthetic spectra from the
Munari and Tomasella (1999b), Munari and Castelli (2000) and Castelli and
Munari (2001) databases.

The radial velocities that we have measured are listed in Table~3. They are
those obtained via cross-correlation. Manual measurements of individual
lines has provided pretty similar results, only less accurate.

\section{Hipparcos/Tycho photometry}

The input photometric data for the three program stars are those provided by
the Hipparcos/Tycho mission. Those for V505~Per and OO~Peg have been taken
directly from the mission final databases on CD-ROMs, while for
V570~Per the data (which were not previously published/distributed given
their poorer quality) have been obtained upon request from ESA.

The Hipparcos/Tycho mission collected data in three photometric bands (cf.
Figure~1). The Hipparcos $H_P$ band is the most accurate of the three (cf.
Table~2) and resembles white light measurements. The Tycho $B_T$ and $V_T$ bands are
similar to the Johnson equivalent bands, as indicated by the transformation
equation provided in the explanatory notes of the Hipparcos/Tycho Catalogue:
\begin{eqnarray}
    V_J & = & V_T \ - \ 0.090 \times (B-T)_T \\
(B-V)_J & = & 0.850 \times (B-V)_T
\end{eqnarray}

\section{Modeling}

The WD98 code used for the analyses is the successor program to  WD95
described by Kallrath et al. (1998), the philosophy behind which was
described in Kallrath and Milone (1999). WD98 includes the version of the
Wilson-Devinney code distributed by Wilson (1998), namely the options of
having ($a$) time rather than phase as input data, ($b$) a square root
limb-darkening law (in addition to the

\clearpage

\noindent
linear and logarithmic options of
WD95), ($c$) atmospheric scattering effects in modeling the light curve,
($d$) adjustment of the period and epoch within the Wilson-Devinney code,
and ($e$) adjusting a linear change in period and in the argument of
periastron. WD98 permits iterative operation, and variable damping factor as
the solution is approached, to minimize the effects of correlations among
the parameters, and retains Kurucz atmosphere models for the stellar
components. We use logarithmic limb-darkening coefficients for the passbands
computed from Kurucz models by W. VanHamme, and the desktop interpolation
program of D. Terrell.

\begin{table*}[!Ht]
\caption[]{Modeling solutions. The uncertainties are formal mean 
standard errors to the solution. The last four rows gives the standard
deviation of the observed points  from the derived orbital solution.} 
\centerline{\psfig{file=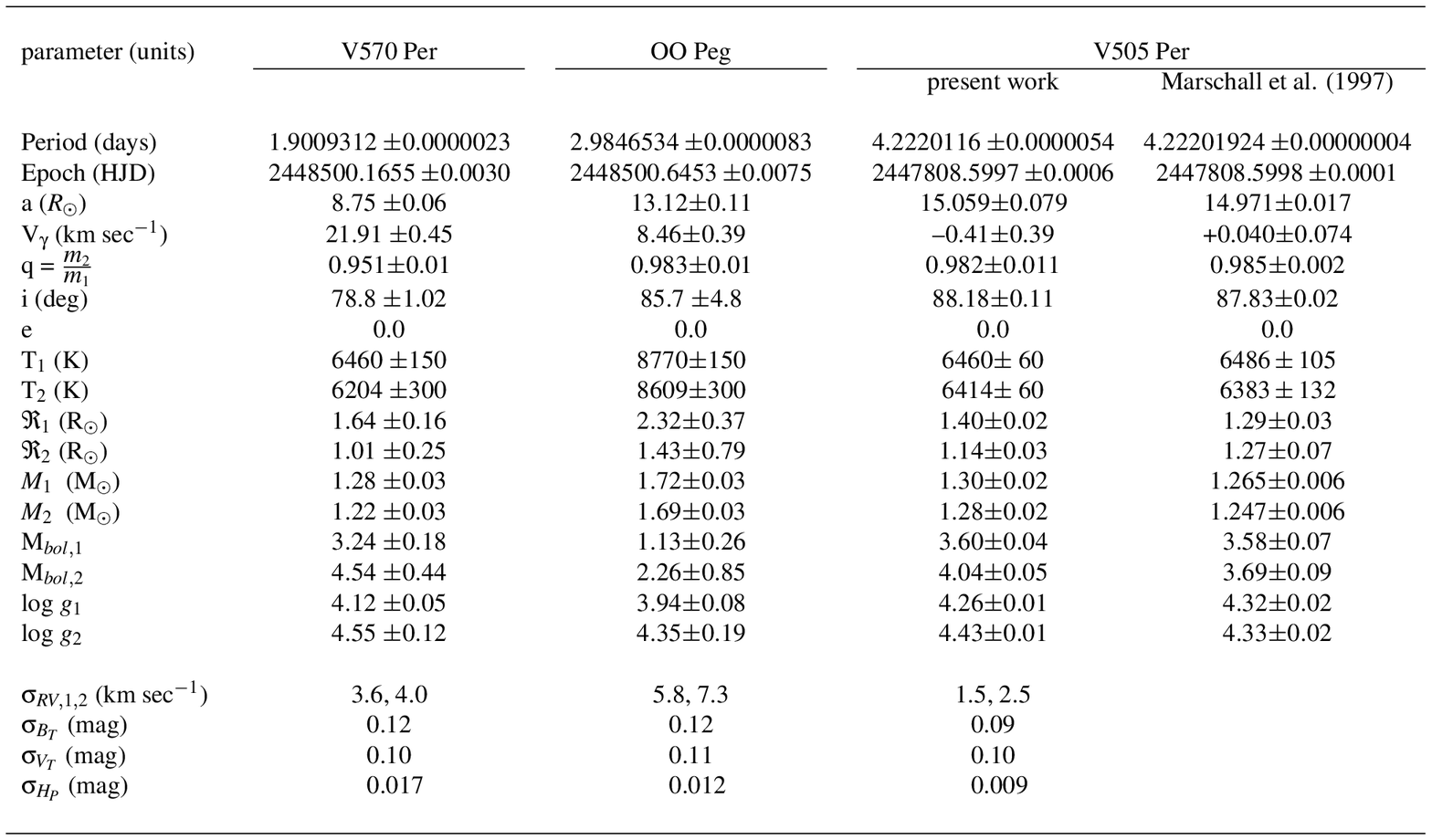,width=17cm}}
\end{table*}

For all the three stars investigated in this study the following steps were
taken. The photometric data were plotted and the mean of the maxima used to
normalize the light of the system. The initial epoch and period were taken
from the Hipparcos solution (listed in the Hipparcos Catalogue), the
temperature of the hotter star from the spectral classification, the
temperature of the cooler star from the ratio of the depth of the minima,
initial potentials set arbitrarily so that the stars did not exceed their
critical surfaces (in modeling contact and over-contact systems, this
procedure is changed, of course). Initially, all runs were carried out with
linear limb-darkening and single-pass reflection treatment. The $H_P, V_T,
B_T$ photometric data and the radial velocity curves for both components
were modeled together. The passband-specific flux ratio files (containing
the ratio of the atmosphere to blackbody fluxes) were created from Kurucz
atmosphere models by C.R. Stagg.

The parameters that were adjusted initially were the semi-major axis ($a$),
baricentric radial velocity ($V_{\gamma }$), inclination ($i$), temperature of
star 2 ($T_{2}$), modified Kopal potentials ($\Omega _{1,2}$), mass ratio
($q=m_{2}/m_{1}$), epoch, period ($P$), and relative luminosity in
each passband of star 1 (L$_{1_{\lambda }}$), where star 2 is the component
closer to us at the part of the light curve designated as phase 0.0 .

After the computer-generated corrections for the epoch and period fell below
their standard errors by at least $50\%$, these parameters were no longer
adjusted in subsequent run, to improve the precision of the remaining
parameters. Note that the potentials of both stars were permitted to vary
(thus the stellar sizes are constrained by the eclipses alone).

\section{Results}

For all program stars the solutions well agreed inside the uncertainties
with null eccentricity. We then fixed $e=0.0$ and rerun the solutions.

As the initial temperature for the primary (hotter) star we adopted the
one corresponding to the spectral type (from Table~1) according to 
Popper (1980). No reddening correction was adopted for these nearby 
binary systems (missing also accurate spectral classifications of both 
components). 

For all the program stars there is no discernible sign of departure from
spherical symmetry for either component: $r_{pole}$ = $r_{point}$ =
$r_{side}$ = $r_{back}$ . The program stars are therefore examples of well
detached eclipsing systems.

The goal of this paper is to evaluate the potentials of GAIA observations on
eclipsing binaries and not to discuss into details the solution obtained for
each program star. To this aim we will refrain from listing and discussing
unnecessary minor details about the modeling solutions. Moreover, we used
all available photometric data to the aim of simulating {\sl automatic}
treatment of GAIA data; it is however clear from Figures~3, 4 and 5 that
using {\sl only} $H_P$ data (ignoring $B_T$ and $V_T$ ones) would have
neatly reduced the errors on some of the stellar parameters in Table~4, in
particular the radii and the derived quantities (like bolometric magnitudes
and surface gravities). Therefore, the software that will actually deal with
GAIA data should be smart enough to exclude from the analysis the whole 
lightcurves adding only noise to the solution.

\subsection{V505 Persei}

\begin{figure}[t]
\centerline{\psfig{file=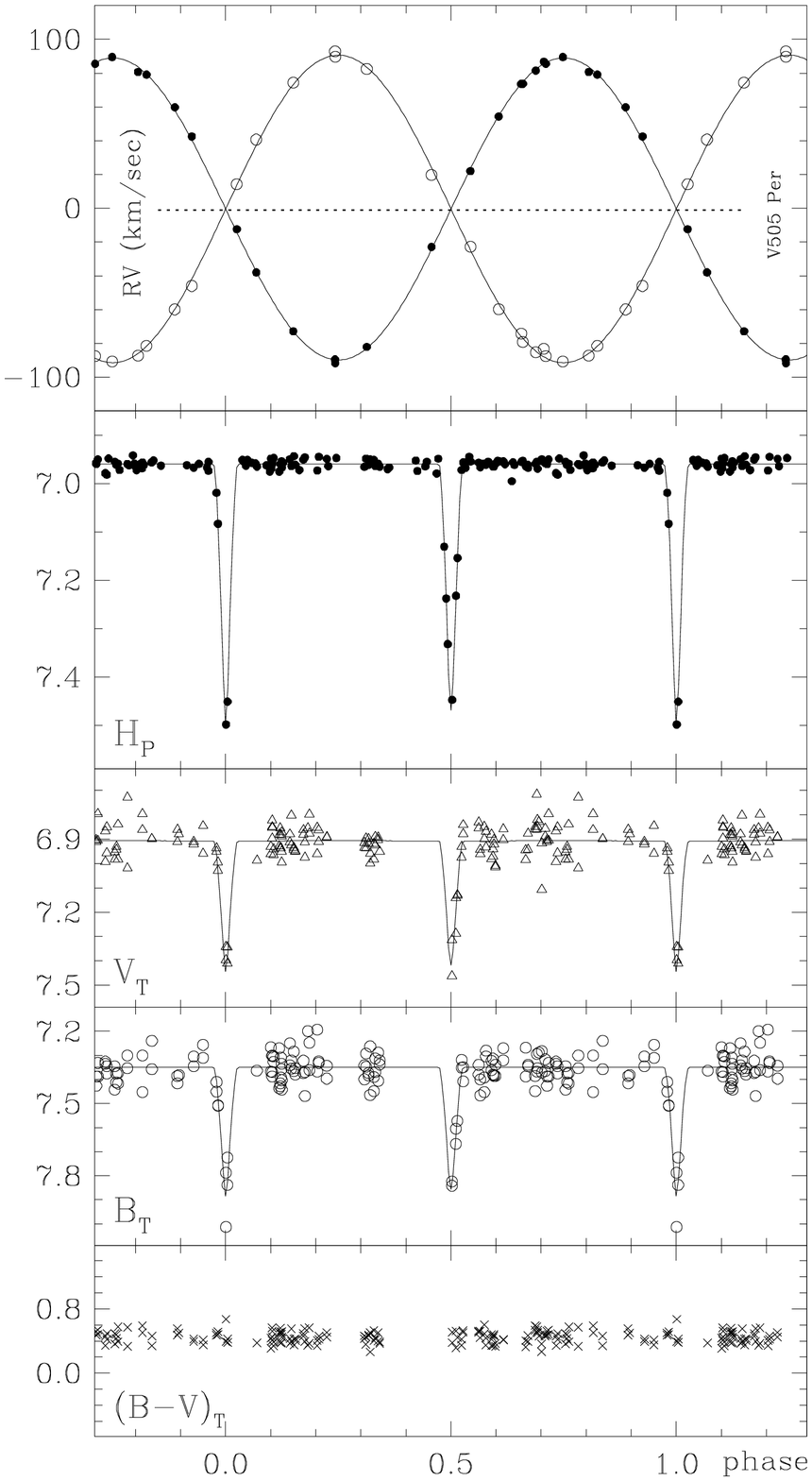,width=8.9cm}}
\caption[]{Hipparcos $H_P$ and Tycho $V_T, B_T, (B-V)_T$ lightcurves of V505~Per
folded onto the period P = 4.2220116 days. The lines represent the
solution given in Table~4.}
\end{figure}

\begin{figure}[t]
\centerline{\psfig{file=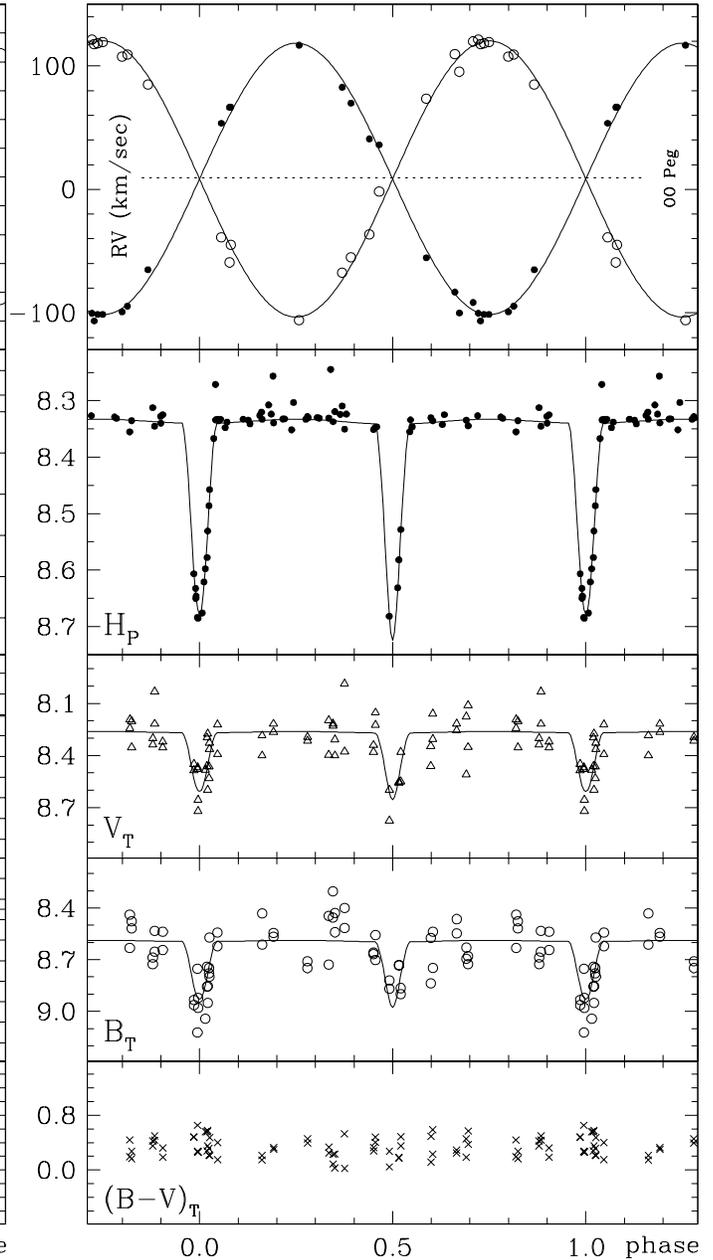,width=8.9cm}}
\caption[]{Hipparcos $H_P$ and Tycho $V_T, B_T, (B-V)_T$ lightcurves of OO~Peg
folded onto a period P = 2.9846534 days. The lines represent the
solution given in Table~4.}
\end{figure}

\begin{figure}[t]
\centerline{\psfig{file=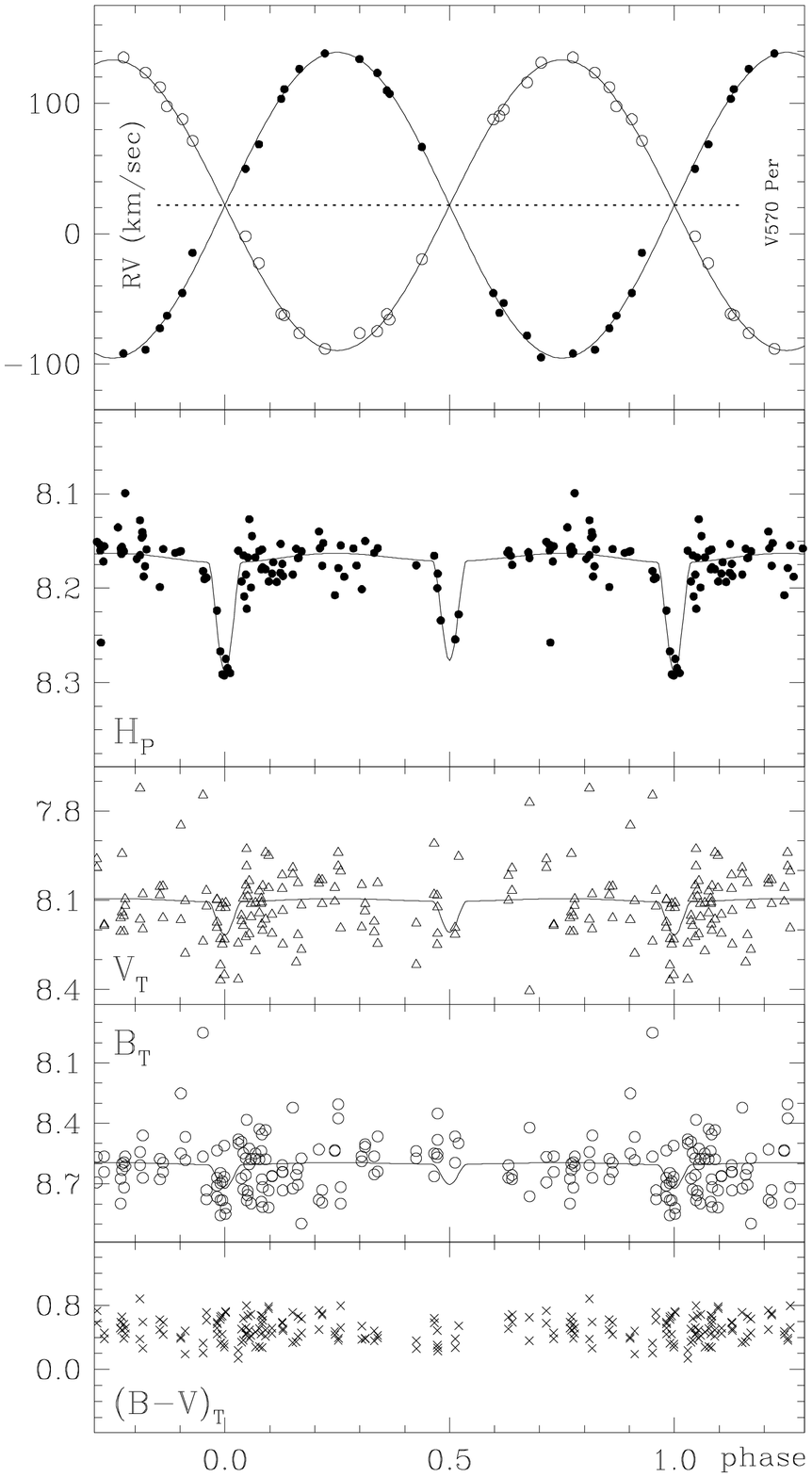,width=8.9cm}}
\caption[]{Hipparcos $H_P$ and Tycho $V_T, B_T, (B-V)_T$ lightcurves of V570~Per
folded onto a period P = 1.9009312 days. The lines represent the
solution given in Table~4.}
\end{figure}

The derived parameters for V505~Per are listed in Table 4, where they are
compared to the Marshall et al. (1997) solution which should be correct to
about 1\% or better, and thus serves at the ``{\sl true, intrinsic}" set of
values to compare with.

Marshall et al. used 63 spectra secured at
R~=~$\lambda/\bigtriangleup\lambda$~=~43,000 compared to our only 20 at
R~=~$\lambda/\bigtriangleup\lambda$~=~17,000. This of course reflects into 
an intrinsicly higher precision of each observation, a firmer radial velocity
orbital solution and a lower scatter of points along the computed curve
for the Marshall et al. work. Nevertheless, determinations of individual
masses and semi-major axis coincide inside the errors.

The surface temperatures of the two components are constrained to $\pm$60 K
using the Hipparcos/Tycho three band photometry. GAIA will obtain
photometric data on $\sim$11 bands, carefully placed and shaped to
enhance the sensitivity, diagnostic capabilities and disentangling abilities
to temperature, luminosity, metallicity and gravity overlapping effects. It
is therefore quite safe to assume that the error on the temperature estimate
that GAIA could achieve on eclipsing binaries similar to V505~Per will be
significantly better than the already good $\pm$60 K allowed by
Hipparcos/Tycho data.

The other parameters are satisfactorily determined, usually coincident
inside the reported formal errors with the Marshall et al. solution.

The largest discrepandy between our GAIA-like solution and Marshall et al.
work is - of course - in the determination of the stellar radii which is
strongly dependent on the branches of the minima in the lightcurve. The
Marshall et al. lightcurve has eclipses mapped by about 400 points of
excellent quality per band, while only 10 Hipparcos $H_P$ and 18 Tycho
$B_T$, $V_T$ measurements cover the eclipses. Even if the {\sl formal} error
to the solution is smaller in our case, this is no doubt the result of the
very small number statistics. The difference in the radii between our and
Marshall et al. solutions is 8.5\% for the primary and 11\% for the
secondary. Looking at the lightcurve in Figure 3 not a single $H_P$
observation falls on the rising branch of the primary eclipse. No wonder
then that a difference of the order of 10\% exist in the radius
determinations ! Just doubling the number of Hipparcos observations (and
thus statistically doubling the eclipse coverage) would have resulted in a
determination of absolute radii of the components of V505~Per accurate to
3-4\%.

\subsection{OO Pegasi}

OO~Peg is similar to V505~Per (well detached system, similar period, equal
depth eclipses), but observationally more difficult: it is fainter (by 1.4
mag) and its early-A spectral type makes the Ca~II lines weaker and the
spectra dominated by broad Paschen lines. In fact, mean standard errors in
Table~2 for photometry and radial velocities are twice those of V505~Per.

Contrary to V505~per and V570~Per, we intentionally did not pay attention
to the orbital phase when we observed OO~Peg, in the attempt to simulate odd
phase coverage by GAIA. This choice generated the quite uneven distribution
of RV points in Figure~4, with just a single observation around the first
quadrature.

In spite of such less favorable photometric and spectroscopic observing
conditions, the solution for OO~Peg in Table~4 looks good anyway. The error
on the semi-major axis is 0.8\% (0.5\% for V505~Per), on the temperature of
the secondary 3\% (0.9\%), on masses of both components 1.1\% (1.5\%), on
mass ratio 1.0\% (1.0\%), on systemic velocity 0.4 km sec$^{-1}$ (0.4 km
sec$^{-1}$).  The much poorer and noisier eclipse sampling of OO~Peg in
comparison to V505~Per manifests itself in the determination of radii and
inclination: the error on the radius of the primary is 16\% (8\%) and for
the secondary 55\% (11\%), while the error for inclination is 4.8 degrees
(0.5 degrees).

\subsection{V570 Persei}

With V570~Per we explored a system very similar to V505~Per, but with twice
larger errors (cf. Table~2) affecting both photometry and spectroscopy {\sl
and} showing much less pronounced eclipses. Indeed, they are visible only in
the $H_P$ lightcurve (cf. Figure~5). As for V505~Per, we planned the
spectroscopic observations to obtain a good coverage at all orbital phases.

The system parameters mainly linked to radial velocities are constrained
well enough: the error of the semi-major axis is 0.7\% (0.5\% for V505~Per),
of masses of both components 2.4\% (1.5\%), of mass ratio 1.1\% (1.0\%), of
systemic velocity 0.45 km sec$^{-1}$ (0.4 km sec$^{-1}$).  Determination of
radii and inclination obviously suffer from poorer sampling and larger noise
of photometric data: the error of the radius of the primary is 10\% for
V570~Per (8\% for V505~Per) and for the secondary 25\% (11\%), while for
inclination the error is 1.0 degrees (0.5 degrees). The temperature of the
secondary star is constrained within 4.5\% (0.9\% for V505~Per).

\section{Discussion}

\begin{table}[!t]
\tabcolsep 0.08truecm
\caption{Comparison between the Hipparcos distances and those derived from the 
parameters of the modeling solution in Table~4.}
\begin{tabular}{lcccc}
\hline
          &&               &             &                  \\
          &&\multicolumn{3}{c}{distance (pc)}               \\ \cline{3-5}
          &&Hipparcos      & this paper  &   Marshall et al.\\
          &&               &             &                  \\
V505 Per  &&  66$\pm$4     &  ~~59$\pm$4 &    62$\pm$3      \\ 
V570 Per  &&117$\pm$14     &   108$\pm$6 &                  \\
OO Peg    &&445$^{304}_{840}$ 
                           & ~\,320$\pm$38&                  \\
          &&               &             &                  \\
\hline
\end{tabular}
\end{table}

A detailed discussion of expected GAIA performances will necessarily be
postponed to the end of the present series, when enough stars will be
investigated to cover the span of spectral types and kinds of interaction
(detached, semi-detached, contact). This will provide us with a
representative sample of the huge zoo of eclipsing binary stars within the
observational capabilities of the coming ESA Cornerstone~6 mission.

Nevertheless, a few brief statements are in order to comment on the
promising results on well detached, double-lined eclipsing binaries of A, F
spectral types of a few days orbital period obtained here.

\subsection{Masses, Orbital Separations, Baricentric Velocities}

In this paper we used a mean of 23 spectra per star, spread over a wide
range of S/N.  (cf. Table~2). Orbital separations and individual masses are
constrained to $\sim$0.7\% and $\sim$1.5\%, respectively.  Thus the selected
$\bigtriangleup \lambda =$250 \AA\ wavelength range between 8490 and 8740
\AA\ is working fine for spectral types A and F, even though it was
optimized for cooler stars (the vast majority of field stars in the range of
magnitude of interest to GAIA are G and K stars). GAIA will record the
majority of its spectra at S/N lower than obtained for this paper, but this
will be compensated by a much larger number of spectra per star, typically
150. It is therefore quite possible that a sizeable fraction of the well
detached, double-lined eclipsing binaries of A-F spectral type observed by
GAIA will have their orbital separation and masses of both components
determined to an accuracy of $\sim$1\% provided that the spectral dispersion
and resolution will be not too far from 0.25 \AA/pix and 0.5 \AA,
respectively.

The baricentric velocities are constrained to better than 0.5 km sec$^{-1}$,
which is fine in comparison to the accuracy of GAIA tangential motions, and
also adequate to resolve the internal kinematics of stellar aggregates
(clusters, associations, etc.)

\subsection{Period, Epoch, Eccentricity}

Periods, epochs and eccentricities can be derived to a high level of
accuracy as this investigation has proved and the 5 years of planned mission
lifetime for GAIA suggests.

For the program stars we adopted $e=0.0$ because modeling excluded any
eccentricity larger than 0.01, which can be taken as a rough indication of
the accuracy to which eccentricity could be determined by GAIA observations.

Periods for the three program stars have been determined to such an accuracy
that it would take $\bigtriangleup t \sim$170 years (on the average) to
bring the ephemeris out of phase for more than 10\%\ of the orbital period.
The 5 years of GAIA operation (compared to the time span of 3.36 years for
Hipparcos data or 1.38 years for the spectral data used here), should expand
the 10\% accuracy horizon of the ephemeris to $\bigtriangleup t \sim$250
years.

\subsection{Inclination, radii}

These parameters severely depend upon the way photometry maps the eclipse
phases.

Hipparcos has collected about 110 measurements per star. If eclipses last
for 1/10 or less of the orbital period, one has to expect 10 or less
photometric points to be distributed over the eclipse phases.

GAIA will have three lines of view (at $\sim$120 deg). If the same set
of core photometric bands is present in {\sl all} the three channels (a feature
yet to be optimized in the currently baselined GAIA design), the number of
collected photometric points will proportionally increase with great benefit
not only to the study of eclipsing binaries but also to variable stars
in general.

\subsection{Temperature, surface gravity}

These two parameters are quite dependent too on GAIA photometry.  However,
they can be determined in a completely independent way.

Effective temperatures and surface gravities are in fact two of the basic
output of spectral synthesis analysis (e.g.\ based on Kurucz model
atmospheres). The 0.25 \AA/pix dispersion and 0.5 \AA\ resolution GAIA
spectra used here are quite adequate to derive temperatures to $\pm$30~K and
gravities to 0.1 dex (cf. Munari and Castelli 2000). The analysis of GAIA
spectra of eclipsing binaries (mainly those obtained at quadratures when
spectra of both components are better observed) would therefore provide
these quantities directly.

The effective temperature and surface gravity determined directly from GAIA
spectra could then be used to check {\sl a posteriori} the results of
modeling or used as inputs in the modeling process to decrease the number of
variables.

\subsection{Comparison with Hipparcos Distances}

A natural check about the goodness of the orbital solution and modeling is
to compare the distance to the program stars computed from the modeling
parameters in Table~4 with the trigonometric parallaxes measured by
Hipparcos. This is done in Table~5, together with the distance to V505~Per
from the modeling of Marshall et al. (1997).

The high expected performances of GAIA observations are confirmed: ($a$) the
modeling of photometric lightcurves and radial velocity curves for the three
program stars constrains the distance to a better accuracy than Hipparcos
trigonometric parallax, and ($b$) modeling distances and Hipparcos
parallaxes agree inside the respective errors.

As for the effective temperature and surface gravity, trigonometric
parallaxes determined directly by GAIA (at an average 200$\times$ better
precision than Hipparcos) could then be used to check {\sl a posteriori} the
results of modeling or used as an input value in the modeling process
itself.

\subsection{Prospects for GAIA}

The unique characteristic of GAIA as a single mission which combines
astrometry, spectroscopy and photometry appears to offer special advantages
compared to the classical ground-based approach to eclipsing binaries. The
possibility to use effective temperatures, surface gravities and
trigonometric parallaxes as fixed input values to the modeling of
lightcurves and radial velocities should allow more confident derivation of
other parameters, both stimulating a new modeling approach to eclipsing
binaries and also granting lower errors to the solution.

By probing all constituents of the Galaxy (halo, disk, bulge) and detecting
eclipsing binaries there, GAIA can foster an epochal leap in our knowledge
of the basic stellar parameters and how they vary with mass, age and
chemical abundance. This will be true not only on statistical grounds (given
the huge number of observable eclipsing systems) but also for an object by
object approach.  This paper has in fact shown that an accuracy of 1\% in
the derived stellar parameters can be achieved by GAIA on a fraction of the
detached eclipsing binaries observed during the mission.

\acknowledgements{
Generous allocation of observing time with the Asiago telescopes has been
vital to the present project as it has been the financial support by the
Osservatorio Astronomico di Padova and by the Italian Space Agency (via
grants to CISAS, University of Padova). It is a pleasure to thank PierLuigi
Bernacca for his support; Panos Niarchos for his careful reading and
commenting upon the manuscript; Bob Wilson for making available his further
improved Wilson-Devinney program and for advice on its usage; Walter Van
Hamme, Dirk Terrell, and Chris Stagg for contributing to the effectiveness
of the WD98 and wd98k93 packages. The financial support of the Italian Space
Agency (ASI), of NSERC and University of Calgary Research Grant Committee
and the Slovenian Ministry for Research and Technology are kindly
acknowledged.}

\end{document}